\documentclass[a4paper]{JHEP3}
\usepackage{amsfonts,amsmath,amssymb}
\newcommand{\be}{\begin{equation}}
\newcommand{\ee}{\end{equation}}

\newcommand{\bea}{\begin{eqnarray}}
\newcommand{\eea}{\end{eqnarray}}

\newcommand{\eda}{\bar{\eta}}

\newcommand{\tta}{\tilde{\eta}}
\newcommand{\sP}{/\!\!\!\partial}
\newcommand{\sF}{\: /\!\!\!\! F}
\newcommand{\nn}{\nonumber}
\newcommand{\dd}{{d}}
\newcommand{\ra}{\rightarrow}
\title{1/4-BPS M-theory bubbles with $SO(3)\times SO(4)$ symmetry}
\author{Hyojoong Kim and Kyung Kiu Kim \\
Department of Physics and Astronomy, \\
Seoul National University, Seoul 151-747, Korea \\
E-mail: \email{kimhj76@phya.snu.ac.kr, kkkeagle@phya.snu.ac.kr}}
\author{Nakwoo Kim \\
Department of Physics and Research Institute of Basic Science, \\
Kyung Hee University, Seoul 130-701, Korea \\
E-mail: \email{nkim@khu.ac.kr}}
\date{}
\abstract{In this paper we generalize the work of Lin, Lunin and
Maldacena on the classification of 1/2-BPS M-theory solutions to a
specific class of 1/4-BPS configurations. We are interested in the
solutions of 11 dimensional supergravity with $SO(3)\times SO(4)$
symmetry,
 and it is shown that such solutions are constructed over a
 one-parameter familiy of 4 dimensional almost Calabi-Yau spaces.
 Through analytic continuations we can obtain M-theory solutions
 having $AdS_2\times S^3$ or $AdS_3\times S^2$ factors. It is shown
 that our result is equivalent to the $AdS$ solutions
 which have been recently
 reported as the near-horizon geometry of M2 or M5-branes
 wrapped on 2 or 4-cycles in Calabi-Yau threefolds.
 We also discuss the hierarchy of M-theory bubbles with
different number of supersymmetries. }

\begin{document}
%
%
\section{Introduction}




We have recently witnessed that a systematic analysis of
supersymmetric solutions in supergravity theories which utilises the
existence of a Killing spinor can lead to a remarkable insight into
string theory and strongly coupled gauge theory via the
gauge/gravity correspondence \cite{Maldacena:1997re}. Especially in
\cite{Lin:2004nb}, the authors consider 1/2-BPS fluctuations of
maximally supersymmetric AdS solutions in type IIB supergravity and
find that the entire set of regular solutions can be matched with
the phase space of one-dimensional free fermions. This is in good
harmony with the dual field theory, ${\cal N}=4, D=4$ super
Yang-Mills model: in the 1/2-BPS sector it is reduced to  a
Hermitian matrix model whose eigenvalues can be treated as free
fermions when one takes into account the Van der Monde determinant.

It is then natural to ask whether we can also identify the gauge dynamics for
less supersymmetric operators from the geometric constraints imposed by
unbroken supersymmetry on the supergravity side. While the 1/2-BPS
solutions are equipped with $SO(4)\times SO(4)$ which results in $S^3\times S^3$
part in the 10 dimensional metric, 1/4-BPS operators have $SO(4)\times SO(2)$
symmetry which implies that the solutions should have a $S^3\times S^1$ factor.
Supersymmetric solutions of type IIB supergravity with such isometries have
been studied in \cite{Donos:2006iy,Donos:2006ms}.
One can also consider 1/8-BPS solutions which have just
a single $S^3$ factor in the metric \cite{Kim:2005ez}, and
it can be shown that the solution is constructed over
a 6 dimensional K\"ahler space obeying a type of non-linear Laplace equation
for the Ricci tensor.
See \cite{Gava:2006pu} for the study on a different class of 1/8-BPS solutions,
and \cite{Chen:2007du} for a unified viewpoint
and a systematic analysis of supersymmetric regular
solutions and the identification of
the dual operators to 1/2, 1/4 and 1/8-BPS solutions.

It is also interesting to apply this program to M-theory.
In \cite{Lin:2004nb} the
authors considered the 1/2-BPS
fluctuations, or bubbles, of M-theory as well and showed that the
supergravity equations are reduced to a 3 dimensional continuous Toda equation.
It is expected that this
particular Toda equation is responsible for the dynamics of 1/2-BPS operators
of the superconformal field theory defined on M2 or M5-branes, although
in this case we do not have a perturbative description of the dual conformal
field theory and it is not clear how to {\it derive} the Toda system from
the field theory. See \cite{Bak:2005ef,Ganjali:2005cr} for discussions on the solutions of the
Toda equation and their interpretations as giant gravitons.

One can try to determine the dual geometry for less-supersymmetric
M-theory bubbles.
1/8-BPS solutions with an $S^2$ factor, or $AdS_2$ when analytically
continued, has been
studied already in \cite{Kim:2006qu} and the resulting BPS system satisfies,
surprisingly enough, exactly the same
equation - now defined in 8 dimensions - which governs $S^3$ bubbles
of IIB theory.
A natural interpretation of such configurations
is that they are dual to BPS operators which are Lorentz singlet and
holomorphic in $SU(4)\subset SO(8)$ R-symmetry of the M2-brane theory.

We are interested in 1/4-BPS bubbles of M-theory in this paper.
If we consider the 6 dimensional field theory of M5-branes with $(2,0)$
supersymmetry and restrict ourselves to BPS operators which are
Lorentz-singlet but holomorphic in $SU(2)\subset SO(5)$ R-symmetry,
 the dual geometry should carry an $SO(6)$ symmetry
which lead to an $S^5$-factor in the metric. A related problem of supersymmetric
$AdS_5$ solutions in M-theory has been addressed in \cite{Gauntlett:2004zh}
and the local geometry of the
corresponding
bubble solutions are obtained simply through analytic continuations.
Although it will be very interesting to study the bubble solutions in detail
and identify the dual operators, in this paper we restrict ourselves to the other class of 1/4-BPS M-theory bubbles.
In M2-brane field theory, if a given Lorentz-singlet operator saturates the
BPS bound and is written as a holomorphic combination of two chiral multiplets,
it should be invariant under $SO(3)\times SO(4)$ symmetry so the dual geometry should contain $S^2\times S^3$.
We take this as our starting point and analyse how the supersymmetry
helps us determine the local form of the solutions, filling the gap between the
1/2-BPS bubbles of \cite{Lin:2004nb} and the 1/8-BPS solutions of
\cite{Kim:2006qu}. There exist a number of papers which explore the
AdS/CFT relation using the supergravity backgrounds  for specific M-brane configurations as duals to interesting field theory objects such as Wilson loops,
defect conformal field theories etc. See for instance
\cite{Lin:2005nh, Lunin:2006xr, Lunin:2007ab}.

Once we establish the $S^2\times S^3$ solutions, it is straightforward
to obtain $AdS_2\times S^3$ or $AdS_3\times S^2$ via a series of analytic
continuations. They are interpreted more naturally as the near-horizon
geometry of (wrapped) M2 or M5-branes with some extra isometries in the
transverse space. Such configurations have been already analysed
using the supersymmetry condition of brane probes in the Calabi-Yau
threefolds, by
\cite{Gauntlett:2006ux,MacConamhna:2006nb}.
We will show that our results indeed agree with the wrapped brane
solutions.

Sec. 2 serves as the main part of this article. We first fix our
convention and derive the 6 dimensional Killing spinor equations in
Sec.2.1. We then analyse the algebraic and differential equations
of spinor bilinears
to determine the geometry of the solutions in Sec.2.2.
In Sec. 3 we discuss how one can obtain the Wick-rotated versions
$AdS_2\times S^3$ and $AdS_3\times S^2$ through analytic continuations,
and show they are equivalent to the results of \cite{Gauntlett:2006ux,MacConamhna:2006nb}.
In Sec.4 and Sec. 5 we discuss how our solutions can be related to 1/2-BPS or 1/8-BPS M-theory
bubbles from the literature. We conclude in Sec. 6.

\section{$S^2\times S^3$ ansatz and the local form of the supersymmetric solutions}
\subsection{The metric ansatz and the Killing spinor equations in $D=6$}
In this paper, we aim to study supersymmetric solutions in 11 dimensional
 supergravity with $SO(3)\times SO(4)$ isometry which
are dual to 1/4-BPS operators  of the dual conformal field theory in 3 or 6 dimensions.  We thus assume that the spacetime metric should contain
$S^2\times S^3$.  More specifically, our ansatz is
\bea
\dd s^2_{11}
&=& e^{2A} \dd s^2_{S^2} + e^{2B} \dd s^2_{S^3}+ g_{\mu\nu} \dd x^\mu \dd x^\nu
,
\\
G &=& F \wedge {\rm Vol}_{S^2} ,
\eea
where $ds^2_{S^2}$ and $ds^2_{S^3}$ represent the metric of
the round sphere with radius 1 in the appropriate dimensionality.
We dimensionally reduce the four-form field strength $G=dC$ to have
a 6 dimensional gauge field $F$. Since electric(magnetic) configurations of
$G$ are associated to M2(M5)-branes, in our setting M2-branes are
wrapped on $S^2$ and M5-branes contain the $S^3$ as part of the worldvolume.

We adopt the standard convention for the 11 dimensional
supergravity with the lagrangian density
\be
{\cal L} = R *1 - \frac{1}{2} G \wedge *G - \frac{1}{6} C\wedge G\wedge G ,
\label{11action}
\ee
and the supersymmetry transformation for the gravitino is given as
\be
\delta \psi_M
=
\nabla_M \epsilon
+ \frac{1}{288}
\left(
\Gamma_M^{\;\;\;M_1 \cdots M_4} - 8 \delta_M^{M_1} \Gamma^{M_2M_3M_4}
\right)
G_{M_1 \cdots M_4}
\epsilon ,
\ee
with the spinorial parameter $\epsilon$ which should obey the Majorana condition.
$\Gamma_M$ represents the 11 dimensional gamma matrices satisfying
\be
\{ \Gamma_M , \Gamma_N \} = 2 g_{MN} ,
\ee
where $g_{MN}$ is the 11 dimensional metric tensor and $M,N=0,1,\cdots, 10$.

Above ansatz can be understood as the dimensional reduction of 11 dimensional
supergravity theory on (unsquashed) $S^2\times S^3$, and we expect to have an effective action in
6 dimensions,
which has the metric, two scalar fields $A,B$, and a two-form field strength $F$
as the dynamical fields. It is worth noting here that in our ansatz the
cubic Wess-Zumino term in (\ref{11action})
has no effect, so from the form-field equation and
the Bianchi identify for 11 dimensional field we know $F$ should satisfy simply
\bea
\dd F &=& 0 , \\
\dd \left( e^{2A-3B} *_6F\right) &=& 0  .
\label{f_eq}
\eea

We need to choose a gamma matrix basis which respects the dimensional
split we have introduced, to derive 6 dimensional Killing spinor equations
from the 11 dimensional one. Our convention is, in Minkowski spacetime,
\bea
\Gamma_{a} &=& \sigma_{a} \otimes  1 \otimes 1 , \quad a=1,2\nn \\
\Gamma_{\alpha} &=& \sigma_3 \otimes \sigma_{\alpha} \otimes \gamma_{7} , \quad \alpha = 1,2,3 \nn \\
\Gamma_{\mu}  &=& \sigma_3 \otimes 1 \otimes \gamma_{\mu}, \quad \mu= 0,1,\ldots,5 .
\eea
where $\sigma$ are the Pauli matrices. For simplicity we will choose the basis where the 6 dimensional gamma
matrices $\gamma_\mu$ and $\gamma_7$ are all antisymmetric.

We can decompose an 11 dimensional Killing spinor as an expansion over
the Killing spinors on $S^2, S^3$, i.e.
\be
\epsilon = \sum_i \left( \zeta_i \otimes \chi_i \otimes \eta_i + c.c. \right) ,
\ee
where $\zeta (\chi )$ is a $2(3)$ dimensional spinor, and
$\eta$ is the Killing spinor of the 6 dimensional system we are
interested in.
On the spheres $S^2$ and $S^3$, the Killing spinor should be
conformally parallel, which means
\bea
\overline{\nabla}_a \zeta &=& \pm \frac{1}{2} \sigma_a \sigma_3 \zeta ,
\label{ks2}
\\
\overline{\nabla}_\alpha \chi &=& \pm \frac{i}{2} \sigma_\alpha \chi ,
\label{ks3}
\eea
where $\overline{\nabla}$ denotes the covariant derivative on the sphere
with unit radius. For definiteness let us choose the positive sign in the above
relations for $\zeta, \chi$. One can then derive the following 6 dimensional
Killing spinor equations from $\delta \psi_M=0$:
\bea
\left[
\sP A - \frac{i}{6} e^{-2A} \sF + e^{-A}
\right]
\eta &=& 0 ,
\label{K1}
\\
 \left[
\sP B + \frac{i}{12} e^{-2A} \sF + i e^{-B} \gamma_7
\right]
\eta &=& 0 ,
\label{K2}
\\
\nabla_\mu \eta  - \frac{i}{48} e^{-2A} \gamma_\mu \sF \eta + \frac{i}{16} e^{-2A}
\sF \gamma^\mu \eta &=& 0 .
\label{K3}
\eea

A comment is in order on different sign choices in (\ref{ks2}) and (\ref{ks3}) and the
number of supersymmetries of our ansatz.  The Killing spinors on the sphere should come in some irreducible representations of the isometry group. They make a doublet of $SU(2)$ for $S^2$, and $(2,1)\oplus (1,2)$ of $SU(2)\times SU(2)$ for $S^3$.
For each of them, we expect to have a nontrivial solution to the 6 dimensional Killing spinor equation given above, so we should have 8 real solutions due to the
Majorana condition in $D=11$. Our ansatz thus should provide 1/4-BPS configurations in general.


\subsection{Spinor bilinears and their properties}
Let us now introduce the differential forms which are defined as
spinor bilinears. We first consider tensors whose components are given
as $\eda\gamma_{\mu_1 \cdots \mu_n}\eta$. Our convention goes as follows:
\bea
C &=& i \eda \eta,
\\
D &=& \eda \gamma_7 \eta  ,
\\
K_\mu &=& \eda \gamma_\mu \eta ,
\\
L_\mu &=& \eda \gamma_\mu \gamma_7 \eta ,
\\
Y_{\mu\nu} &=& \eda \gamma_{\mu\nu} \eta ,
\\
Y'_{\mu\nu} &=&i  \eda \gamma_{\mu\nu} \gamma_7 \eta ,
\\
Z_{\mu\nu\lambda} &=& i \eda \gamma_{\mu\nu\lambda} \eta ,
\\
W_{\mu\nu\lambda\rho} &=& i \eda\gamma_{\mu\nu\lambda\rho} \eta .
\eea
Note that they are all real-valued. One can of course also define additional
tensors such as $Z'_{\mu\nu\lambda} = i\eda\gamma_{\mu\nu\lambda}\gamma_7\eta$,
but it is Poincare dual to $Z$. We will see shortly that the $D=11$ solution is built upon a $D=4$ Kahler space, so it is essentially the lower-rank tensors up
to 2-forms which are needed to specify the local geometry of supersymmetric
solutions.

Due to antisymmetry of $\gamma_\mu$,
tensors such as $\eta^T\gamma_7\eta, \eta^T\gamma_\mu\eta,
\eta^T\gamma_\mu\gamma_7\eta, \eta^T\gamma_{\mu\nu}\eta$ vanish identically.
We can easily see $\eta^T\eta=0$ for nontrivial solutions from (\ref{K1}) or
(\ref{K2}). We are thus left with the following tensors,
\bea
\omega_{\mu\nu} &=& \eta^T \gamma_{\mu\nu} \gamma_7 \eta ,
\\
\phi_{\mu\nu\lambda} &=& \eta^T \gamma_{\mu\nu\lambda} \eta ,
\\
\psi_{\mu\nu\lambda\rho} &=& \eta^T \gamma_{\mu\nu\lambda\rho} \eta ,
\eea
which are in general complex-valued.

Now we are ready to
study the geometry of supersymmetric backgrounds using
the existence of Killing spinors. We exploit
the differential and algebraic constraints
from the Killing equations and Fierz identities to identify
the local form of the supersymmetric solutions.

Let us start with the scalars $C,D$.
If we multiply $\eda$ to (\ref{K1}) and ({\ref{K2}),
\be
e^{-A} C = 2 e^{-B} D = -\frac{1}{6} e^{-2A} \eda \sF \eta .
\label{scalar_id}
\ee
Furthermore, when we take the derivative of $C$, we get
\bea
\partial_\mu C
&=&
\frac{1}{12} e^{-2A} \eda [\sF , \gamma_\mu ] \eta
\\
&=& \partial_\mu A \, C .
\eea
So, we fix the normalization of $\eta$ and set
\be
C = e^A , \quad\quad D = e^B/2 .
\ee
From now on
we will make use of these relations whenever we come across $C,D$.

Now let us turn to the vectors.
From the Fierz identity one can prove that
\bea
K \cdot L &=& 0 ,
\label{f1}
\\
K^2 + L^2 &=& 0 .
\label{f2}
\eea
and $K$ is time-like, whereas $L$ is space-like. One can also prove that
in general
\be
|\eta^T\eta|^2 = \frac{1}{2} (L^2- K^2) -(C^2 + D^2) .
\ee
But since $\eta^T\eta=0$, we have
\be
L^2 = - K^2 = e^{2A} + \frac{e^{2B}}{4} .
\label{f_t}
\ee
Readers are referred to Appendix for details on Fierz rearrangement identities
in 6 dimensions.

From the Killing spinor equations, it is straightforward to verify that
\be
\nabla_{ ( \mu } K_{\nu )} = 0 ,
\ee
which implies $K$ defines a Killing vector.  And we can also see from
the Killing spinor equation that
the isometry of the metric associated with $K$ is actually a symmetry
of the whole solution. The Lie derivatives of scalar fields $A,B$ and
gauge field $F$ all vanish. As a one-form,
its exterior derivative is given as
\be
d(e^A K) = F + Y .
\label{d_k}
\ee

For the other vector field $L$, from the algebraic relations we can derive
\bea
L_\mu &\equiv& \eda \gamma_\mu \gamma_7 \eta
\nn
\\
&=&
\frac{1}{2} e^{-B} \partial_\mu ( e^{A+2B} ) ,
\label{defL}
\eea
and from the differential Killing spinor equation (\ref{K2}),
\be
\nabla_\mu L_\nu
=
-\frac{i}{48} e^{-2A} \eda ( \sF \gamma_{\mu}\gamma_\nu + \gamma_{\nu}\gamma_{\mu} \sF )
\gamma_7 \eta
+ \frac{i}{16} e^{-2A} \eda
( \gamma_\mu \sF \gamma_\nu + \gamma_\nu \sF \gamma_\mu )
\gamma_7 \eta ,
\ee
leading to a significant requirement:
\be
\nabla \cdot L = 0 ,
\label{divL}
\ee
while the exterior derivative satisfies
$d(e^{-A/2} L)=0$, which is consistent with (\ref{defL}).

Now let us try to specify the 6 dimensional metric using the information
we have collected so far. From the time-like Killing vector $K$, we
introduce a time-like coordinate $t$ and  set $K=\partial_t$. With $L$
we define a space-like coordinate as $y=e^{A/2+B}$ and set
$L = e^{A/2}dy$. It will be convenient to define a scalar $\zeta$ as
\be
\sinh \zeta = \frac{1}{2} y e^{-3A/2} ,
\ee
to simplify the following discussions.
In this coordinate system, we may write down the 6 dimensional metric as
\be
ds^2_6 =
-e^{2A} \cosh^2 \zeta  \, (dt + \sum_{i=1}^4 V_i dx^i )^2
+ \frac{e^{-A}}{\cosh^2\zeta} dy^2 + e^{-A} \sum_{i,j=1}^4
h_{ij} dx^i dx^j .
\label{met1}
\ee
Note that
we have introduced a warp factor $e^{-A}$ for the 4 dimensional space ${\cal M}_4$ with metric $h_{ij}$ for later convenience.

The problem is now effectively reduced to 4 dimensions. When we introduce
the gauge potential as $F=dB$, and expand
\be
B = B_t dt + B_i dx^i + B_y dy ,
\ee
we have the following unknown functions in 4 dimensions.
\begin{enumerate}
\item metric $h_{ij}$
\item scalars $A, B_t,B_y$
\item vectors $V_i, B_i$
\end{enumerate}
and they all depend on the 5 dimensional spatial coordinates $y, x^i$ in general.

Equipped with the local form of the metric, we are now in a position to
choose an orthonormal frame. We set
\bea
e^{0} &=& e^A \cosh \zeta (dt +V) ,
\\
e^{5} &=& e^{-A/2} \mbox{sech} \,\zeta dy ,
\\
e^{i} &=& e^{-A/2} \hat{e}^i ,  \quad \quad i=1,2,3,4.
\eea
where $\hat{e}^i$ is an orthonormal frame of the 4 dimensional metric
$h_{ij}$.

Our system in general preserves 1/4 supersymmetry of the 11 dimensional
supergravity, and the relevant projection rules can be best expressed using
the orthonormal frame given above.
From the algebraic Killing spinor equations we can eliminate the term with $\sF$ to obtain
\be
\left(
\gamma_{\hat{5}} \cosh \zeta
+ \sinh\zeta
+ i \gamma_{\hat{7}}
\right)
\eta = 0 ,
\label{proj1}
\ee
where the gamma matrices with hatted indices are defined in the tangent space.
We can simplify (\ref{proj1}) in terms of
$\tilde{\eta} = e^{\zeta/2 \gamma_{\hat{5}}} \eta$ and obtain
\be
( 1 + i \gamma_{\hat{5}} \gamma_{\hat{7}} ) \tta = 0 .
\label{pr1}
\ee
Considering $L=e^A\cosh\zeta e^5$, one can find the other projection condition
\be
(1 - i \gamma_{\hat{0}} ) \tta
= 0 ,
\label{pr2}
\ee
and the normalization of $\tta$,
\be
{\tta}^\dagger \tta = e^A .
\label{nor}
\ee
The projection rules imply that $\tta$ is a chiral spinor in ${\cal M}_4$.
As it is well known, an invariant Weyl spinor in $2n$-dimensional space defines an
$SU(n)$-structure, and the intrinsic torsion can be inferred from the derivatives
of the invariant tensors which are constructed as spinor bilinears
\cite{Gurrieri:2002wz, LopesCardoso:2002hd}.

The 4 dimensional $SU(2)$-invariant tensors are included in the 6
dimensional spinor bilinears we have constructed, and we only need to see
how the 6 dimensional tensors are decomposed into 4 dimensions. One can
either directly evaluate each component of the tensors using (\ref{pr1}), (\ref{pr2}) and (\ref{nor}), or make use of the
appropriate Fierz
identities.
Recall first it is our convention that
\bea
K &=& - e^{2A} \cosh^2 \zeta \,  (\dd t+V) ,
\label{dk}
\\
L &=& e^{A/2} \dd y .
\label{dl}
\eea
From (\ref{con12a}) and (\ref{con12b}),
\be
Y= \frac{1}{2} (\dd t+V)\wedge y \,  \dd y + J ,
\label{dY}
\ee
where $J=\frac{1}{2} J_{ij} dx^i \wedge dx^j$ is a 2-form in 4 dimensions which
may have a nontrivial dependence on $y$. The higher-rank tensors turn
out to be products of one- and two-forms given above.
One can also easily see that the 3-form
$Z$ can be written as
\bea
Z &=& - e^{-A} K \wedge Y
\label{de_z}
\\
&=&  e^{A} \cosh^2\zeta \, (\dd t+V) \wedge J .
\eea
and the 4-form $W$ is
\bea
W &=&
-\frac{1}{2}e^{-A} Y\wedge Y
\label{de_w}
\\
&=& -\frac{1}{2} e^{-A} J\wedge J
-  (\dd t+V) \wedge y\, \dd y \wedge J .
\eea
One can also consider the complex-valued 2-form $\omega$ and find it is
a 2-form purely
in ${\cal M}_4$ as one can readily see from (\ref{oc1}) and (\ref{oc2}).
In addition to that, we have
\bea
\phi &=& -e^{-A/2}
\left( \frac{y}{2} (dt+V) + i \frac{dy}{\cosh^2\zeta} \right)
\wedge \omega ,
\label{de_phi}
\\
\psi &=& e^{A/2} (dt+V) \wedge dy \wedge \omega .
\label{de_psi}
\eea

From the direct evaluation or the normalization properties such
as (\ref{y2}) and (\ref{o2}),
we see that $J$ can be used to define an almost complex structure with metric $h_{ij}$,
and $\Omega=({\rm sech} \, \zeta ) \cdot \omega$ provides the properly normalized $(2,0)$-form, satisfying
\be
\Omega \wedge J = 0 , \quad\quad
\mathrm{Vol} ({\cal M}_4) = \frac{1}{4}
\Omega \wedge \overline{\Omega} = \frac{1}{2} J\wedge J  .
\ee

The 6 dimensional derivatives can be decomposed with respect to our coordinate choice,
so we can write
\be
\dd  = \dd_4 + \dd y \wedge \partial_y +
\dd t \wedge \partial_t .
\ee
We now resume the computation of exterior derivatives for our spinor bilinears.
Again employing the algebraic and differential Killing spinor equation, one easily
obtains
\be
\dd Y= 0 .
\label{de_Y}
\ee
When rephrased in 4 dimensional language, it implies
\bea
\dd_4 J &=& 0 ,
\label{d_j}
\\
\partial_y J &=& -\frac{y}{2} {\dd}_4 V ,
\\
\partial_t J &=& 0 .
\eea
One can also see that
\be
{\dd}\omega = 0 ,
\label{d_om}
\ee
which implies
\bea
{\dd}_4 (\cosh \zeta \cdot \Omega) &=& 0 ,
\label{d_o}
\\
\partial_y (\cosh\zeta \cdot \Omega)&=& 0  ,
\\
\partial_t (\cosh\zeta \cdot \Omega)  &=& 0 .
\eea
The 4 dimensional derivatives of the $SU(2)$ tensors (\ref{d_j}) and (\ref{d_o})
determine the $SU(2)$ structure and the intrinsic torsion of ${\cal M}_4$.
From the fact that the $(1,1)$ form $J$ is $d_4$-closed, and their
exists a $(2,0)$ form $\cosh\zeta \, \Omega$ which is also closed,
we conclude that ${\cal M}_4$ is almost Calabi-Yau \cite{Joyce:2001xt}.
One notes that (\ref{d_o}) can be expressed as
\be
\dd_4 \Omega = i P \wedge \Omega ,
\ee
with
\be
P = \frac{3}{2} \tanh^2 \zeta (J \cdot \dd A) .
\ee
As it is well-known, $P$ is the Ricci potential whose exterior
derivative gives the Ricci form $\Re = dP$.

For higher-rank tensors, after similar manipulations we obtain
\bea
\dd(e^{2A}Z) &=& 2 e^A W - F\wedge Y ,
\\
\dd(e^A W ) &=&  0 ,
\\
d(e^{A/2} \phi ) &=& \frac{1}{2} e^{-A/2} \psi .
\label{d_phi}
\eea
We can check that these equations automatically hold once we
demand the supersymmetry conditions given in previous paragraphs.
It is basically because these higher-rank tensors are expressed as
exterior products of 1 and 2-forms, as given in
(\ref{de_z}),(\ref{de_w}),(\ref{de_phi}), and (\ref{de_psi}),
 and do not pose genuinely new
invariant tensors.

The gauge field $F$ can be determined by (\ref{d_k}) once the geometric data
and scalar field $A$ are given. When decomposed into 4 dimensions we have
\be
F = - \dd (e^{3A}) \wedge (\dd t +V)
+ e^{3A} \cosh^2 \zeta \partial_y V \wedge \dd y + \hat{F} ,
\ee
where $\hat{F}$ represents the 4 dimensional part of $F$ and is given as
\be
\hat{F} = - J - e^{3A} \cosh^2 \zeta \dd_4 V .
\ee
At this stage, we can make use of the algebraic Killing equation (\ref{K2})
to derive various constraints on $F$. In particular, we consider
(\ref{f_0}-\ref{f_2}) and find the following relations,
\bea
(\dd_4 V)_+ &=& \frac{\partial_y \left( y^2 e^{-A} \right)}{y\cosh^2\zeta}  J ,
\\
\partial_y V  &=& -\frac{3 \sinh^2\zeta }{2\cosh^4\zeta} J \cdot  \dd A .
\eea
We now see that the Ricci potential $P$ can be written more
succinctly as
\be
P = - \frac{1}{\cosh^2\zeta}  \partial_y V ,
\ee
and when we take $\dd_4$,
\be
y \partial_y \left(  \frac{1}{y} \partial_y J \right)
 =
 \dd_4
 \left(
 J \cdot \dd \, \mathrm{sech}^2 \, \zeta
 \right) .
 \label{me_4}
 \ee
 This equation can be considered as a higher dimensional analogue of
 the Toda equation for the 1/2-BPS fluctuations considered in \cite{Lin:2004nb}.
 1/4-BPS bubbles of IIB supergravity satisfies a very similar differential
 equation, see (58) of \cite{Donos:2006iy}.

 We are now in a position to check whether our supersymmetric configurations
 described so far automatically satisfy the classical field equations.
 As well-established by now, supersymmetry requirements combined with the Bianchi
 identity and the form-field equations imply that the Einstein equation is
 satisfied, unless the Killing spinor is null \cite{Gauntlett:2002fz}.
 For the solutions of our interest in this paper,
 $K^2=-L^2<0$ so the Killing spinor is not null. From the
 equations (\ref{d_k}) and (\ref{dY}) it follows that
 \be
 \dd F = 0 .
 \ee
 Now let us check the field equation (\ref{f_eq}). Among the various
 supersymmetry requirement conditions, we use (\ref{d_k}), (\ref{al_4}),
 (\ref{de_z}) and (\ref{de_w})
 to obtain an expression for $*F$ in terms of the geometric data including $A$.
 \be
 *F = e^{-B} Y\wedge Y + 2 e^{3A-B} \dd ( e^{-2A} K\wedge Y ) .
 \ee
 Now  we can check (\ref{f_eq}) using the 4 dimensional decompositions of $K,Y$ given in (\ref{dk}), (\ref{de_Y}). It is straightforward to
 see that it vanishes provided $\partial_y (\cosh^2\zeta J\wedge J)=0$. But this
 is a consequence of (\ref{divL}), or equivalently (\ref{d_o}). So we have now
 eastablished that the equations of motion are satisfied for our
 supersymmetric configurations.


\section{Analytic continuation to $AdS_2\times S^3$ and $AdS_3\times S^2$}
We have so far considered a specific class of supersymmetric solutions in
D=11 supergravity: configurations with an $S^2\times S^3$ factor. If one is
interested in similar problems, for instance M-theory
 solutions with $AdS_2\times S^3$,
obviously
the same technique can be used to first derive the 6 dimensional Killing spinor
equations and then study the local form of the metric and form-fields constrained by
unbroken supersymmetries. But since we are interested  in solutions containing a product of maximally
symmetric spaces with the same dimensionalities, we can simply take
multiple analytic continuations to transform our results on $S^2\times S^3$
to $AdS_2\times S^3$ or $AdS_3\times S^2$. Actually, such new solutions might have
even more significance in general. $AdS_ 2\times S^3$ is the near-horizon
geometry of 5 dimensional black holes, so the general form of the metric
can be very useful in the systematic study of 5 dimensional supersymmetric
black holes embedded in  11 dimensional supergravity. $AdS_3\times S^2$
solutions are potentially dual to 2 dimensional supersymmetric conformal field
theory whose R-symmetry has an $SU(2)$ factor.

Alternatively, one can also interpret the $AdS$ solutions as a near-horizon
limit of M2 or M5-branes. If one recalls that in our ansatz we have
dimensionally reduced
the 4-form field of the 11 dimensional supergravity on $S^2$, one can
easily conclude that the $AdS_2\times S^3$ solutions are purely M2-brane
configurations while the $AdS_3\times S^2$ solutions are composed purely
of M5-branes. Since we have 1/4-BPS solutions, we can consider
either intersection of two M-branes, or M-branes wrapped on supersymmetric
cycles of Calabi-Yau 3 manifolds, to obtain the desired solutions.

In fact, $AdS$ solutions as near-horizon limits of wrapped M-branes
have been systematically studied recently, first for M5-branes
in \cite{Gauntlett:2006ux} and also for M2-branes in
\cite{MacConamhna:2006nb}. The authors used the fact that the Killing spinors
of the supergravity configurations should obey the same projection rule
required for the probe brane action, and found the local form of the solutions
in an efficient way using the calibration conditions.
 $AdS_3\times S^2$ solutions are given in (6.8-6.15)
of \cite{Gauntlett:2006ux}, and $AdS_2\times S^3$ solutions given
in (4.12-4.19) of \cite{MacConamhna:2006nb}. One can check that these
$AdS$ solutions are exactly the same as our solution, albeit written
in different variables. Here we briefly sketch how to establish the equivalence
of \cite{Gauntlett:2006ux} and our results. A similar relation can be also
easily found with $AdS_2\times S^3$ solutions of \cite{MacConamhna:2006nb}.
It is useful first to note that
the triplet of almost complex structures $J^1,J^2,J^3$ which describe
the 4-dimensional base space in \cite{Gauntlett:2006ux} are translated
in our convention as $J^1\ra e^{-A}J , J^2+iJ^3 \ra  e^{-A}\Omega$.
Now it is straightforward to check that (6.10) and (6.11) in \cite{Gauntlett:2006ux}
correspond to (\ref{d_om}). Similarly, (6.12) of \cite{Gauntlett:2006ux} is
equivalent to (\ref{de_Y}). In particular, when we complexify (6.13) and (6.14),
the resulting equation is equivalent to (\ref{d_phi}).

In the rest of this subsection we illustrate how one can analytically continue
$S^2\times S^3$ solutions to obtain $AdS$ solutions, and write the form of
the metric for easier reference.
By analytic continuation we mean we set all the coordinates of the round sphere
to pure imaginary. For instance, start with the 2-sphere with metric
\be
\dd s^2 ( S^2 ) = \dd\theta^2 + \sin^2 \theta \dd \phi^2 ,
\ee
and through the reparametrization  $\theta=i\rho, \phi=i\tau$, the metric becomes
\bea
\dd s^2 &=& -\dd\rho^2 + \sinh^2 \rho \, \dd \tau^2
\\
&=& - \dd s^2 (AdS_2 ) .
\eea
To fix the overall sign of the metric, we further take the re-definition
$e^{2A}\rightarrow - e^{2A}$ but leave $e^{2B}$ invariant, or
$y^2 \rightarrow i y^2$. In particular, now the metric can be written as
\be
\dd s^2 =
e^{2A} \dd s^2_{\mathrm{AdS}_2} + y^2 e^{-A} \dd s^2_{\mathrm{S}^3}
+ e^{2A} \cos^2\zeta (\dd\psi+V)^2 + \frac{e^{-A}}{\cos^2\zeta} \dd y^2
+ e^{-A} h_{ij} \dd x^i \dd x^j ,
\ee
where we introduced a space-like coordinate $\psi$ by setting $t\rightarrow \psi$.
$\zeta$ is defined as
\be
\sin \zeta = \frac{1}{2} y e^{-3A/2} ,
\ee
so for consistency the range of $y$ is restricted to satisfy $\sin^2\zeta \le 1$, unlike
the $S^2\times S^3$ solutions.

It is also straightforward to consider $AdS_3\times S^2$. The metric can be
written as
\be
\dd s^2 =
e^{2A} \dd s^2_{\mathrm{S}^2} + y^2 e^{-A} \dd s^2_{\mathrm{AdS}_3}
+ \frac{y^2e^{-A}}{4} \cos^2\xi (\dd\psi+V)^2 + \frac{4e^{2A}}{y^2\cos^2\xi} \dd y^2
+ e^{-A} h_{ij} \dd x^i \dd x^j ,
\ee
with
\be
\sin \xi = \frac{2}{y} e^{3A/2} .
\ee
\section{Examples and the identification of K\"ahler spaces}
The most prominent examples of M-theory solution with $S^2\times S^3$ are
certainly the maximally supersymmetric configurations $AdS_4 \times S^7$ and $AdS_7\times S^4$.
For definiteness here we consider $AdS_4\times S^7$ and re-write the
metric in a way compatible with our results in this paper. The
other case of $AdS_7\times S^4$ can be treated in a similar way.

Let us first start with the 11 dimensional metric, which can be written as follows
to make $SO(3)\times SO(4)$ isometry manifest.
\bea
ds^2_{11} &=& R^2 \left[ d\rho^2 - \cosh^2\rho \, dt^2 + \sinh^2\rho \, ds^2_2
+ 4 ( d\theta^2 + \sin^2\theta ds^2_3 + \cos^2\theta d\tilde{s}^2_3 ) \right] .
\eea
Let us choose $ds^2_2$ and $ds^2_3$ as the part corresponding to our $S^2$ and $S^3$.
Obviously, we can identify as
\bea
e^{2A} &=& \sinh^2 \rho ,
\\
e^{2B} &=& 4\sin^2 \theta ,
\eea
so
\bea
\sinh\zeta &=& \frac{\sin\theta}{\sinh\rho} .
\eea
In order to identify the 4 dimensional locally K\"ahler space, we split the metric of
$S^3$ using the left-invariant forms of SU(2).
\bea
\dd \tilde{s}^2_3 &=& \frac{1}{4} \left( \sigma_1^2 + \sigma_2^2 + \sigma^2_3
\right)
\\
&=& \frac{1}{4} \left[ (\dd\psi + \alpha)^2 + \sigma_1^2 + \sigma^2_2 \right]
,
\eea
where $\dd\alpha = \sigma_1\wedge \sigma_2$.
Now if we take the  re-parametrization $\psi \rightarrow \psi+t$ the 6 dimensional
part of the metric becomes
\bea
\dd s^2_6 &=& - (\sinh^2\rho+\sin^2\theta) \left[\dd t - \frac{\cos^2\theta}{\sinh^2\rho+\sin^2\theta} \sigma_3 \right]^2
\nonumber \\
&&+\frac{\cosh^2\rho\cos^2\theta}{\sinh^2\rho+\sin^2\theta}
\sigma_3^2 + \dd\rho^2 + 4 \dd\theta^2 + \cos^2\theta (\sigma^2_1+\sigma^2_2) .
\label{47a}
\eea
First of all we can now see the identification
\be
V = - \frac{\cos^2\theta}{\sinh^2\rho+\sin^2\theta} \sigma_3 .
\ee
In order to identify the 4 dimensional K\"ahler part which is transverse to $K,L$
vectors, it is required to compute $\frac{e^{-A}}{\cosh^2\zeta} dy^2$ part of
(\ref{met1}) and subtract it from (\ref{47a}). Upon the change of  coordinates
\bea
y &=& 2 \sqrt{\sinh\rho} \sin\theta ,
\label{inv_1}
\\
v &=& 2\sqrt{\cosh\rho} \cos\theta ,
\label{inv_2}
\eea
it is straightforward to check
\be
\dd\rho^2 + 4 \dd\theta^2 = \frac{\sinh\rho}{\sinh^2\rho+\sin^2\theta} \dd y^2
+ \frac{\cosh\rho}{\sinh^2\rho+\sin^2\theta} \dd v^2 ,
\ee
where $\rho,\theta$ are now treated as functions of $y,v$ implicitly through
the inversion of
(\ref{inv_1}), (\ref{inv_2}).
Now we can write down the metric of ${\cal M}_4$ which is expected to have
a locally K\"ahler structure.
\be
\dd s^2_{{\cal M}_4} = \sinh\rho
\left[
\frac{\cosh\rho}{\sinh^2\rho+\sin^2\theta} \dd v^2
+\frac{\cosh^2\rho\cosh^2\theta}{\sinh^2\rho+\sin^2\theta} \sigma^2_3
+ \cos^2\theta (\sigma^2_1 + \sigma^2_2 )
\right] .
\ee
The conditions on the $SU(2)$-structure of ${\cal M}_4$, such as the equations
which are derived from $\dd Y=\dd\omega=0$, can be verified once we
fix the complex structure, or the K\"ahler form of ${\cal M}_4$. It turns out that
we need to choose
\be
J = \frac{\sinh\rho\cosh^{3/2}\rho\cos\theta}{\sinh^2\rho+\sin^2\theta}
dv\wedge \sigma_3
+ \sinh\rho \cos^2\theta \, \sigma_1\wedge \sigma_2 ,
\ee
then it is straightforward to check that
\bea
\dd_4 J &=& 0 ,
\\
\partial_y J  &=& -\frac{y}{2} \dd_4 V ,
\eea
indeed hold. The rest of the constraints can be also shown to be satisfied.

We next consider the 1/2-BPS bubble solutions of M-theory
obtained in \cite{Lin:2004nb}.
The relevant little group of the supersymmetric states is $SO(3)\times SO(6)$,
which should appear as $S^2\times S^5$ within the dual geometry. The Killing
spinor analysis has been performed in \cite{Lin:2004nb} and we
quote the result here,
\bea
ds^{2}&=& -4e^{2\lambda}(1+\tilde{y}^{2}e^{-6\lambda})
(dt+\tilde{V}_{i}dx^{i})^2 +\frac{e^{-4\lambda}}
{1+\tilde{y}^{2}e^{-6\lambda}}[d\tilde{y}^{2} + e^{D}(dx_{1}^2+dx_{2}^{2})] \nonumber\\
&& + 4e^{2\lambda}ds^2 (S^5) + \tilde{y}^{2}e^{-4\lambda}ds^2 (S^2) ,
\\
G &=& \mathrm{Vol}(S^2) \wedge F ,
\\
e^{-6\lambda} &=& \frac{\partial_y D}{y(1-y\partial_y D)} ,
\\
\tilde{V}_i &=& \frac{1}{2} \epsilon_{ij} \partial_j D ,
\\
F &=& dB_t \wedge (dt+\tilde{V}) + B_t d\tilde{V} + d\hat{B} ,
\\
B_t &=& - 4 \tilde{y}^3 e^{-6\lambda} ,
\\
d\hat{B} &=& 2 *_3 \left[ (y \partial^2_y D + y (\partial_y D)^2 -
\partial_y D ) dy + y \partial_i \partial_y D dx^i \right] .
\eea
The scalar function $D$ satisfies a 3 dimensional version of the
Toda equation
\be
(\partial_1^2+\partial_2^2) D + \partial^2_y e^D = 0 .
\label{toda}
\ee

In order to identify the 4 dimensional almost Calabi-Yau space
as a verification of our result,
we first write $S^5$ as a fibration over $S^3$,
\be
d\Omega_{5}^{2} = d\alpha^2 + \cos^2\alpha d\psi^2 +
 \sin^2\alpha ds^{2} (S^3),
\ee
Obviously one can identify
\bea
e^{2A} &=& \tilde{y}^2 e^{-4 \lambda} , \\
e^{2B} &=& 4 e^{2 \lambda} \sin^2\alpha  .
\eea
In order to identify the 4 dimensional K\"ahler base, we
first shift $\psi \ra \psi +t$, and introduce a new set of
coordinates $(y,v,z_1,z_2)$ from $(\tilde{y}, \alpha, x_1,x_2)$ as
follows
\bea
y &=& 2 \sqrt{\tilde{y}} \sin\alpha , \\
u &=& e^{D/2}\cos\alpha ,\\
z_1 &=& x_1 , \\
z_2 &=& x_2 .
 \eea
Then one can show that the metric tensor becomes
  \bea
ds_{11}^{2}&=&\tilde{y}^{2}e^{-4\lambda}ds^{2} (S^2)+
4 e^{2\lambda} \sin^2\alpha \, ds^2(S^3) \nonumber\\
&-& 4(e^{2\lambda}\sin^2\alpha+\tilde{y}^2 e^{-4\lambda})
\left[dt+\frac{(1+\tilde{y}^2 e^{-6\lambda})\tilde{V}-\cos^2\alpha d\psi}{\sin^2\alpha
+\tilde{y}^2 e^{-6\lambda}} \right]^2 +\frac{\tilde{y} e^{-4\lambda}}{\sin^2\alpha+\tilde{y}^2 e^{-6\lambda}}dy^2\nonumber \\
&+& \tilde{y}^{-1}e^{2\lambda}\Big\{4\tilde{y} \cos^2\alpha
\frac{1+\tilde{y}^2 e^{-6\lambda}}{\sin^2\alpha+\tilde{y}^2 e^{-6\lambda}} (d\psi-\tilde{V})^2
+\frac{\tilde{y}e^{-6\lambda}}
{1+\tilde{y}^{2}e^{-6\lambda}} e^{D}(dz_{1}^2+dz_{2}^{2}) \nonumber\\
&+&4\tilde{y}e^{-D} \frac{1+\tilde{y}^2 e^{-6 \lambda}}{\sin^2\alpha+\tilde{y}^2 e^{-6\lambda}}
\left[du+e^{D/2}\cos\alpha(\tilde{V}_2 dz_1 -\tilde{V}_1 dz_2)\right]^2
\Big\} .
\eea
We choose the K\"ahler form as
\bea
J&=&4\tilde{y}\cos\alpha~ e^{-D/2}\frac{1+\tilde{y}^2 e^{-6 \lambda}}{\sin^2\alpha+\tilde{y}^2 e^{-6\lambda}}
\Big(du+e^{D/2}\cos\alpha(\tilde{V}_2 dz_1 -\tilde{V}_1 dz_2)\Big) \wedge (d\psi-\tilde{V}) \nonumber\\
 &-&\frac{\tilde{y}e^{-6\lambda}}{1+\tilde{y}^{2}e^{-6\lambda}} e^{D} dz_1
 \wedge dz_2 ,
\eea
and one can check that $d_4 J=0$ and $\partial_y J = -\frac{1}{2}y d_4 V$,
using (\ref{toda}). The $(2,0)$-form $\Omega$ is taken as follows,
\bea
\Omega &=& e^{i\psi} \left( 4 \tilde{y}
\frac{\tilde{y} e^{-6 \lambda}}{\sin^2\alpha+\tilde{y}^2 e^{-6\lambda}}
\right)^{1/2} \cdot
\nonumber\\
&&
\Big[
(du+e^{D/2}\cos\alpha(\tilde{V}_2 dz_1 -\tilde{V}_1 dz_2))+
i(d\psi -\tilde{V})\Big]
\wedge (dz_1 - i dz_2) ,
\eea
which satisfies
\bea
d_4 (\cosh\zeta \Omega)=0, \quad\quad \partial_y (\cosh\zeta \Omega) = 0 .
\eea


\section{The relation to 1/8-BPS AdS Bubbles}
\label{5}
Supersymmetric M2-brane configurations with an $AdS_2$ factor in the metric
has been studied in \cite{Kim:2006qu}. It turns out that the 9 dimensional
internal space should take the form of a warped U(1)-fibration over an 8 dimensional
K\"ahler space ${\cal M}_8$. One can also easily translate the results
into the case of solutions with an $S^2$, instead of $AdS_2$,
through analytic continuation. They would in general provide 1/8-BPS
 bubbles of M-theory.
The specific type of solutions with $S^2\times S^3$
we have
studied in this paper can be considered as a special case of such
1/8-BPS solutions. In this section we show how the 1/4-BPS solutions studied in
this paper can be re-written in a way as presented in
\cite{Kim:2006qu,Gauntlett:2006ns}.

Let us first briefly summarize the result of \cite{Kim:2006qu}. One starts with
the following ansatz:
\bea
ds^2 &=& e^{2\bar{A}} \left[ ds^2(S_2) + ds^2(Y_9) \right] ,
\label{s2}
\\
G &=& \mathrm{Vol}(S^2) \wedge F .
\eea
The existence of a nontrivial Killing spinor restricts the local form of the solution
as follows,
\bea
\dd s^2 (Y_9) &=& - (\dd t + P)^2 + e^{-3\bar{A}} \dd s^2 ({\cal M}_8 ) ,
\label{s2_9d}
\\
F &=&
 J + \dd
\left[ e^{4\bar{A}} ( \dd t + P ) \right] .
\eea
${\cal M}_8$ is required to be  K\"ahler with K\"ahler form $J$, and Ricci potential $P$.
The warp factor is also determined purely by the geometric data of
${\cal M}_8$,
\be
e^{-3\bar{A}} = -\frac{1}{2} R .
\ee
The Einstein equation combined with the supersymmetry requirement demands that
the Ricci tensor of ${\cal M}_8$
should satisfy the following equation.
\be
\Box R - \frac{1}{2} R^2 + R_{mn} R^{mn} = 0 .
\label{m_eq}
\ee
One can construct new $AdS_2$ (or $S^2$) solutions in 11 dimensional supergravity
based on a solution of (\ref{m_eq}). Indeed, a countably infinite number of new
$AdS_2$ solutions in M-theory have been obtained in \cite{Gauntlett:2006ns}
using a co-homogeneity 1 solution of (\ref{m_eq}).

Let us now try to identify the 8 dimensional space from the result we obtained
in this paper. Obviously we first identify the two $S^2$'s in
(\ref{met1}) and (\ref{s2}) and set $\bar{A}=A$. We then
write the metric of $S^3$ explicitly using the left-invariant forms of SU(2)
\be
\dd s^2 (S^3) = \frac{1}{4} \left(
\sigma^2_1 + \sigma^2_2 + \sigma^2_3
\right) .
\ee
Now upon a coordinate shift $\sigma_3 \ra \sigma_3 +t$, we can re-arrange the metric (\ref{met1}) into a form found in (\ref{s2}) and (\ref{s2_9d}), and identify
the metric of 8 dimensional K\"ahler base as
\bea
ds^2 ({\cal M}_8) &=&
\mathrm{sech}^2\, \zeta \, dy^2 + \frac{y^2}{4}\cosh^2\zeta \, (\sigma_3-V)^2
+ \frac{y^2}{4} (\sigma^2_1+\sigma^2_2) + \sum_{i,j=1}^4 h_{ij} dx^i dx^j .
\eea
And the Ricci potential is given as
\be
P = V\cosh^2\zeta - \sinh^2\zeta \, \sigma_3 .
\label{rip}
\ee

In order to check the consistency conditions we introduce the K\"ahler form
of ${\cal M}_8$ as
\be
J_8 =
\frac{y}{2} dy \wedge (\sigma_3-V)
+\frac{y^2}{4} \sigma_1\wedge \sigma_2 +
J_4 ,
\ee
where $J_4$ denotes the K\"ahler form of ${\cal M}_4$. One can easily check
$d J_8=0$ using $\partial_y J_4 + \frac{2}{y} \dd_4 V = 0 $. The associated
$(4,0)$-form is given as
\be
\Omega_8 = \left( \frac{y}{2} \mathrm{sech} \zeta dy + i \frac{y^2}{4}\cosh\zeta (\sigma_3-V)
\right) \wedge (\sigma_1+i \sigma_2) \wedge \Omega_4 .
\ee
It is also straightforward to check $d\Omega_8 = i P \wedge \Omega_8$ with
$P$ given as (\ref{rip}), so $dP$ indeed gives the Ricci-form of ${\cal M}_8$.
\section{Discussions}
In this paper we have used the technique of Killing spinor analysis
to determine the geometric constraints imposed by the requirement of
supersymmetry and $SO(3)\times SO(4)$ isometry in M-theory.
The main motivation for this work has been to generalize the AdS bubble
solutions of \cite{Lin:2004nb} to 1/4-BPS solutions. Like other examples of supersymmetric
AdS bubbles reported earlier in
\cite{Donos:2006iy,Donos:2006ms,Kim:2005ez,Kim:2006qu}, it turns out that the 11 dimensional spacetime is based on a
K\"ahler subspace. It is natural to associate this symplectic structure with
the phase space of the gauge field dynamics for the BPS sector.
We have derived a set of partial differential equations
which determines the K\"ahler base space and eventually the 11 dimensional
metric and the gauge field. Technically the partial differential equations
can be derived if one first considers $AdS_2\times S^3$ or $AdS_3\times S^2$
and continue analytically to $S^2\times S^3$ case. The relevant $AdS$ solutions
have been already studied in \cite{Gauntlett:2006ux} and \cite{MacConamhna:2006nb}.
We argued that all of them essentially lead to the same equations, in Sec.3.

Once we have reduced all the equations of motion in 11 dimensions
down to 5 dimensions spanned by $y,x^i$, the next step is to solve
the equations like (\ref{me_4}) and obtain new solutions. We leave this
 task for future publications, and put more emphasis on
the hierarchy of K\"ahler spaces associated with different types of AdS
bubbles. 1/2-BPS bubbles of \cite{Lin:2004nb}, including the maximally
supersymmetric solutions, provide nontrivial solutions of (\ref{me_4}).
In turn, the solutions presented in this paper would automatically satisfy
another highly nontrivial equation (\ref{m_eq}) which
describes the dynamics of  1/8-BPS bubbles.

It is also very important to find the connection of our results
with the dual field theory dynamics. For 1/2-BPS bubbles of IIB theory,
where the field theory is amenable to perturbative analysis since it is
reduced to a hermitian matrix model, there has been considerable
progress in relating the Yang-Mills theory with the semiclassical
treatment of IIB supergravity theory
\cite{Grant:2005qc,Dai:2005hh,Yoneya:2005si,Dhar:2005su,Berenstein:2007wz,Skenderis:2007yb}.
See also \cite{Larjo:2007zu} for analogous discussions on bubbles of
$AdS_3\times S^3$. Together with the insight one earns from
the concrete computations on both sides of the duality in the above
works, we hope that our results on the geometry of supergravity
solutions play an important role in
uncovering the microscopic building block of the
dual conformal field theory on M-branes.

\acknowledgments
The research of N. Kim is supported by the Science Research Center Program of the Korea Science and Engineering Foundation through the Center for Quantum Spacetime (CQUeST) of Sogang University with grant number R11-2005-021.
H. Kim and K. Kim are supported by the Korea Research Foundation Grant funded by the Korean Government
(MOEHRD)(R14-2003-012-01002-0).

\appendix{}
\section{Algebraic relations between the spinor bilinears}
A number of algebraic equations can be derived for the spinor bilinears from
the algebraic Killing equations (\ref{K1}) and (\ref{K2}). We first start with
(\ref{K1}) and take contractions with $\eda$ after multiplying different numbers
of gamma matrices. If we multiply $\eda$ we have
\bea
K^\mu \partial_\mu A &=& 0 ,
\\
F_{\mu\nu} Y^{\mu\nu} + 6 e^{2A}  &=& 0 .
\label{f_0}
\eea
And if we take contractions with $\eda\gamma_\mu$ to get one-form equations
we obtain
\bea
\partial_\mu A + \frac{1}{3} e^{-2A} F_{\mu\nu} K^\nu &=& 0 ,
\\
K_\mu - \frac{1}{6} e^{-A} Z_{\mu\nu\lambda} F^{\nu\lambda}
+ e^A Y_{\mu\nu} \partial^\nu A &=& 0 .
\eea
Similarly the two-form identities are
\bea
F_{\mu\nu} - \frac{e^{-A}}{2} W_{\mu\nu\alpha\beta} F^{\alpha\beta}
+ 3 Y_{\mu\nu} + 3 \left( K_\mu \partial_\nu A - K_\nu \partial_\mu A\right) &=& 0 ,
\label{f_2}
\\
Z_{\mu\nu\alpha} \partial^\alpha A
+ \frac{1}{3} e^{-2A}
\left(
Y_{\mu\alpha} F_{\nu}^{\;\;\alpha} -
Y_{\nu\alpha} F_{\mu}^{\;\;\alpha}
\right)
&=& 0 .
\eea
Let us present a 4-form equation also here which plays a crucial role
when we check the gauge field equation of motion. One multiplies
$\eda\gamma_{\mu\nu\lambda\rho}$ to (\ref{K1}) and find
\be
W - \frac{1}{3} Y\wedge F + \frac{1}{6} e^{-A+B} *F + e^A Z \wedge dA = 0 .
\label{al_4}
\ee

One can also first eliminate $\sF$ in the equation and construct various
spinor bilinear, i.e. start with
\be
\left[
\sP (A+2B) + e^{-A} + 2 i e^{-B} \gamma_7
\right]
\eta = 0 .
\ee
If we multiply $\eda$ from left the real part gives
\be
K^\mu \partial_\mu (A+2B) = 0 ,
\ee
and the imaginary part is
\be
D = {e^{-A+B}}  C / 2 .
\ee

Below we list several of such algebraic relations.
\bea
L^{\mu} \partial_\mu (A+2B) &=& 2 e^{-B} C + e^{-A} D ,
\\
L &=& \frac{e^B}{2} C d(A+2B) ,
\\
Y' &=& - \frac{1}{2} e^{-A+B} Y - e^{-A} K\wedge L ,
\\
Z' &=& + e^A d(A+2B) \wedge Y' ,
\\
W &=& - e^{A-B} * Y + d(A+2B) \wedge Z .
\eea

\section{Fierz identities}
In this section we present a list of useful Fierz rearrangement identities for 6
dimensional commuting spinors. Our Killing spinor system is very similar
to the 1/4-BPS solutions considered in \cite{Donos:2006iy}, and we find the appendix
very useful. Readers are referred to \cite{Donos:2006iy} for more identities and detailed
derivations. In this section
we will repeat some of the derivations in \cite{Donos:2006iy} and rephrase
them in our convention for quick reference and self-sufficiency.
We will also consider
identities involving $\eta^T\eta$.
In particular it will be shown how to derive (\ref{f_t}).

In our convention $\gamma_\mu$ are all antisymmetric and generate
6 dimensional Clifford algebra. The chirality is defined in terms of
\be
\gamma_7 = \gamma_{0123456} ,
\ee
and the positive(negative) chirality part of a spinor $\eta$ is given as
$\eta_\pm = \frac{1}{2}(1\pm\gamma_7)\eta$.

The basic relation for Fierz rearrangement is (see (63) in \cite{Donos:2006iy})
\bea
\eda_1\eta_2 \eda_3\eta_4
&=&
\frac{1}{8}
\left(
\eda_1\eta_4 \eda_3\eta_2
+
\eda_1\gamma_7\eta_4 \eda_3\gamma_7\eta_2 \right)
\nonumber\\
&-&
\frac{1}{16}
\left( \eda_1\gamma_{\mu\nu} \eta_4 \eda_3 \gamma^{\mu\nu} \eta_2
+\eda_1\gamma_{\mu\nu}\gamma_7\eta_4\eda_3\gamma^{\mu\nu}\gamma_7\eta_2
\right)
\nonumber\\
&+&
\frac{1}{8}
\left( \eda_1\gamma_{\mu} \eta_4 \eda_3 \gamma^{\mu} \eta_2
-\eda_1\gamma_{\mu}\gamma_7\eta_4\eda_3\gamma^{\mu}\gamma_7\eta_2
\right)
\nonumber\\
&-&
\frac{1}{96}
\left( \eda_1\gamma_{\mu\nu\lambda} \eta_4 \eda_3 \gamma^{\mu\nu\lambda} \eta_2
-\eda_1\gamma_{\mu\nu\lambda}\gamma_7\eta_4\eda_3\gamma^{\mu\nu\lambda}\gamma_7\eta_2
\right) .
\eea
If we choose
$\eda_1=\eda_\pm\gamma_\mu,\eta_2=\eta_\pm, \eda_3=\eda_\pm$ and $\eta_4=\gamma_\mu\eta_\pm$ one can derive
\be
(K\pm L)^2 = 0 ,
\ee
which in turn implies (\ref{f1}) and (\ref{f2}).

If one uses $\eda_1=\eda_+ ,\eta_2=\eta_-, \eda_3=\eda_-$ and $\eta_4=\eta_+$
we get
\be
C^2+D^2 = \frac{1}{4}(L^2-K^2) + \frac{1}{48} (Z^2 - Z'^2) .
\ee
In order to prove (\ref{f_t}),  one chooses $\eda_1=\eta^T_+,\eta_2=\eta_-,
\eda_3=\eda_-$ and $\eta_4=\gamma_0 \eta^*_+$, to find
\be
|\eta^T\eta|^2 =
\frac{1}{4}(L^2-K^2) - \frac{1}{48} (Z^2-Z'^2) ,
\ee
and as a result we can verify (\ref{f_t}).

It is also desirable to compute the square of two-forms $Y$ and $\omega$.
Choosing $\eda_1=\eda_+ ,\eta_2=\eta_-, \eda_3=\eda_+$ and $\eta_4=\eta_-$,
we get
\bea
-C^2+D^2 &=& \frac{1}{6} (Y^2 - Y'^2 ) ,
\\
C D &=& \frac{1}{6} Y \cdot Y'  ,
\eea
and we also consider $\eda_1=\eta^T_+,\eta_2=\eta_-,
\eda_3=\eda_+$ and $\eta_4=\gamma_0 \eta^*_-$ and find
\be
|\eta^T\eta|^2 =
-\frac{1}{4}(C^2+D^2) + \frac{1}{8} (Y^2+ Y'^2) .
\ee
We can thus conclude, using $\eta^T\eta=0$,
\be
Y^2 = -2 C^2 + 4 D^2 + 4 | \eta^T \eta |^2 ,
\quad\quad Y'^2 = 4 C^2 - 2 D^2 + 4 | \eta^T \eta |^2 ,
\label{y2}
\ee
We can also use $\eda_1=\eta^T_+,\eta_2=\gamma_0\eta_-^*,
\eda_3=\eda_+$ and $\eta_4=\eta_-$ to obtain
\be
| \eta^T\eta |^2 =
\frac{1}{2} \omega \cdot \omega^*  - 4 (C^2 + D^2) .
\label{o2}
\ee

In order to see the decomposition of 6 dimensional tensors in terms of
4 dimensional ones, we need to compute their contractions with $L$ and $K$.
The results are given in (75), (76), (77) and (78) of Donos.  In our notation
they become
\bea
\mathrm{i}_K Y &=& D L \label{con12a}, \\
\mathrm{i}_L Y &=& D K \label{con12b}, \\
\mathrm{i}_K Y' &=& C L,  \\
\mathrm{i}_L Y' &=& C K .
\eea
To compute the contraction of $\omega$ with one-forms we consider
$\eda_1=\eda_\pm\gamma_\mu,\eta_2=\eta_\pm,
\eda_3=\eta^T_\pm$ and $\eta_4=\gamma^\mu\gamma_\nu\eta_\mp$ and find
\be
\eda\gamma_\mu (1\pm\gamma_7) \eta  \cdot \eta^T \gamma^\mu\gamma_\nu(1\mp\gamma_7)\eta = 0 ,
\ee
which leads to
\bea
\mathrm{i}_K \omega &=& \frac{1}{2} (\eta^T\eta) L ,
\label{oc1}
\\
\mathrm{i}_L \omega &=& \frac{1}{2} (\eta^T\eta) K .
\label{oc2}
\eea

\end{document}